# Assessment of the Electromagnetic Behaviour of Servovalve Torque Motor Using Reluctance Network Models

RIBOUT Marion[1,2], ATTAR Batoul[1], HENAUX Carole[3], LLIBRE Jean-François[2] and MESSINE Frédéric[2]

[1] Fluid Actuation & Control Toulouse, Union, France
[2] LAPLACE, Université de Toulouse, CNRS, INPT, UPS, Toulouse, France
[3] IES, Université de Montpellier, Montpellier, France
Email: ribout@laplace.univ-tlse.fr

## ABSTRACT

The torque motor is the most common technology used in electrohydraulic two-stage servovalves to drive the hydraulic pilot stage. As it is a key component in these valves, its performance considerably affects the overall performance of servovalve systems. Modeling accurately the magnetic behavior of the torque motor will help to get a more realistic performance of the servovalve at different operating conditions during the product design and development stage. In this paper, an advanced model of the torque motor is presented. The aim is to integrate it into the overall multi-physics model of the servovalve through a multi-variable function obtained using the proposed modeling tool. The advanced reluctance network model has been developed by progressively considering various magnetic phenomena, including flux leaks between different parts and variations in magnetic induction. This model allows rapid assessment of the magnetic performance at various operating temperatures or using different ferromagnetic materials, without the need for additional analysis in 3D finite element software.

## KEYWORDS

Electromagnetic performance, Reluctance Network, Servovalve, Torque Motor, Magnetic COEnergy, Maxwell stress Tensor, 1D multi-physics model

## I INTRODUCTION

The torque motor is an electromagnetic actuator with limited angle displacement. Its principle is well presented in the Fluid Power literature, as in (Merritt 1967). It consists of two fixed iron parts (Pole Shoes), two permanent magnets and a moving iron part (Armature) supporting two coils fed along the same magnetization axis and separated by the pivot axis, as depicted later in Figure 2. Its performance considerably affects the servovalve performance and, consequently, the overall system performance. Modeling accurately the magnetic behavior of the torque motor will allow a more realistic assessment of the servovalve performance under different operating conditions during the product design and development phase. The 1D multi-physics model is regularly used to rapidly obtain an initial and reliable evaluation of the overall performance of physical systems, as in Simcenter Amesim$^{TM}$, Matlab®/Simulink® and Dymola® modeling environments. In such a model, the torque motor performance can be assessed in different ways. The simplest way is to represent the electromagnetic torque of this motor by a mathematical function, which depends on the angular position of the armature and the consumed current by the coils. For instance, a simple linear function is available in (Merritt 1967) with only two coefficients for current and position effects. These coefficients can be identified according to experimental measurements or to Finite Element Method (FEM) software analysis of the motor performance. This solution is widely used by servovalve manufacturers, as in (Changbin, Chenjun, and Zongxia 2015). However, this approximative model doesn't consider the temperature or the ferromagnetic material effect on the electromagnetic performance.

To get more realistic representation of motor performance, its performance mapping can be incorporated in 1D multi-physics model. This mapping is usually generated, during development phase, by FEM software for a specific ferromagnetic material and at different operating conditions (current, temperature, rotational position, etc.). For example, in a study by (Yan et al. 2017), performance mapping was conducted using JMAG® software to analyze the effect of temperature on the magnet and on the overall performance of the motor.

Even though using FEM software is easy and give good electromagnetic performance estimation, its main drawback is that it is time consuming. Analysis time increases with the motor size, the number of analyzed points and requested accuracy regarding the mesh size. Additionally, it varies according to tested operating conditions or materials (e.g., iron magnetization level). For instance, in the case of a product from our company FACT (Fluid Actuation & Control Toulouse), conducting JMAG® performance analysis may require as long as 12 hours for 7 positions of the armature, analyzed at 3 values of current, and at fixed operating temperature.

Another commonly used method is to use electromagnetic components already available in 1D software, as in Simcenter Amesim$^{TM}$. This software is apparently easy to use, but in fact, it requires a certain level, if not assured, of knowledge in magnetism to reproduce the physical model by a reluctance model that considers different magnetic fluxes paths, especially leakages. This knowledge will be mandatory to select or to criticize the magnetic model embedded in the component icon. For instance, it is noted that in Amesim$^{TM}$, the torque is estimated by considering only the normal component of the magnetic induction in iron and completely neglecting the tangential component. This approach is not correct for the torque motor under study, where the tangential component in iron at some interfaces is at least equal or even bigger than the normal magnetic induction, as it is presented in our previous work (Ribout et al. 2023). Furthermore, it assumes a constant value of magnetic induction in iron parts, which is not correct because of leakage and the effect of tangential flux. Finally, the magnetic reluctance is defined by the length and the cross-section of the flux path, which are unknowns for leakage reluctances.

In this study, a new modeling tool is presented, which allows to assess the torque motor performance in a 1D multi-physics model using a non-linear mathematical function based on an advanced magnetic reluctance model.

This paper is structured to present a state-of-the-art overview examining the current reluctance models for the torque motor in section II. Then, the methodology employed in this research is elaborated upon in section III. Section IV is dedicated to describing the reluctance models development and their performance evaluation. The modeling tool with advanced model is applied to assess the torque variation at different operating conditions in section V. Finally, the paper provides a comprehensive summary of the modeling methodology and the obtained results.

## II STATE OF THE ART

In Fluid Power literature, the torque motor is presented by a simple reluctance network model, as in (Merritt 1967). This simple model does not accurately represent the electromagnetic performance of the motor because of the poor representation of the permanent magnet magnetic behavior and ignoring the various flux leakages. It suggests a simple formula with only two coefficients, as described previously, to evaluate the electromagnetic torque as a function of current and armature position.

In later studies, this reluctance model has been improved by taking into account the correct magnet magnetic behavior and the flux leakage between the both static pole shoes, as in (Urata 2007). The flux leakage coefficient, which is the ratio between leakage flux and useful flux produced by the permanent magnets, has been determined experimentally in this study.

Then, the influence of ferromagnetic materials is considered in the model, as in (Liu and Jiang 2016). Indeed, the effect of ferromagnetic materials saturation cannot be neglected when the magnetic circuit is saturated. In addition, (Zhang et al. 2020) enhanced the reluctance network model by adding a flux leakage reluctance between the two edges of the armature to reduce the effective flux that flowing in the air gap, which makes no physical sense without any saturation in armature. Recently, (Meng et al. 2022) has inserted several flux paths between the pole shoes and the armature, the associated reluctances of each depending on the armature position.

In all these papers, the leakage reluctance is presented by a leakage coefficient that is identified according to experimental measurements or to FEM software analysis results. However, it would be more relevant to consider the real reluctances which depends on the motor geometry rather than the ratio of leakage flux which depends on the operating point.

Furthermore, all previous studies have used the Maxwell stress tensor to calculate the electromagnetic torque of the motor. This method assumes that the tangential component of the magnetic field in the air gap is zero and that only the fluxes passing in the air gaps generate a rotating force on the armature, which may be called into doubt by the occurrence of magnetic leakages.

## III DEVELOPED METHODOLOGY

In this paper, the suggested reluctance network model considers various torque motor reluctances, such as the permanent magnet reluctances, ferromagnetic materials reluctances, and the leakage reluctances. These later mainly presented by three leakage reluctances: the leakage reluctance of permanent magnets, the leakage reluctance between the two stationary pole shoes and the leakage reluctance between the pole shoe and the armature. To be noted, the last leakage reluctance depends on the position of the armature, unlike the other reluctances.

This model will be a part of a proposed modeling tool to predict the motor performance at design and development phase. It generates a mathematical function of electromagnetic torque using the motor different parts dimensions and coils turns number, the used ferromagnetic materials properties, and a single process of performance characterization (via FEM) at fixed temperature. The output function calculates the torque whatever the applied current, the armature position, the operating temperature, and ferromagnetic materials, as described in Figure 1. It can be easily integrated and tested in 1D modeling software.

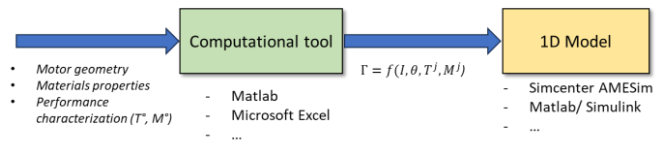

*Figure 1. Inputs/Outputs of proposed modeling tool*

Different incremental reluctance models are analyzed. For each model, the electromagnetic torque has been evaluated using two analytical methods (Maxwell Tensor and COEnergy), as described later (section 4.1). Then, the model that best represents the motor performance observed in FEM software will be implanted in this computational tool to obtain the multi-variable torque function.

In this computational tool, all reluctances of the advanced model, which are expressed as a function of the motor geometry, will be identified using FEM results at only one operating point (one temperature for an infinite relative permeability soft material).

The proposed methodology has many advantages, such as the reduced calculation time, which no more depends on the motor size, contrary to FEM software. The simulation takes about 2min to get the parametrized reluctance network and about 30s to get the electromagnetic performance for a specified temperature and ferromagnetic material. Furthermore, it is easy to be integrated into 1D software, allowing temperature or materials assessment without the need for additional JMAG adjustment, hence saving time. Nevertheless, it depends on the materials state, saturated or not, exactly as any FEM software, where the calculation time can become longer. Indeed, the developed model considers the non-linear pattern of the relative magnetic permeability of high-induction soft ferromagnetic materials. Moreover, the saturation of soft ferromagnetic material can lead to inaccurate estimation of leakage reluctances, which no longer depend exclusively on the motor geometry, as introduced in (Ribout et al. 2023).

In this paper, the study will be carried out for a torque motor already developed by FACT, which is illustrated in Figure 2, see the patent (Ozzello 2021). Furthermore, the Matlab environment is used for calculation, JMAG® as the FEM software.

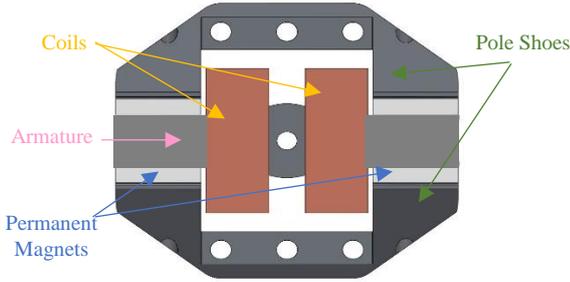

*Figure 2. 2D view of the torque motor under study*

# IV MODELS DEVELOPMENT AND PERFOMANCE EVALUATION

## 4.1 Leakage reluctance models

The torque motor inherently suffers from magnetic flux leakage, as described in (Ribout et al. 2023). In this paper, the effect of leakage between different parts of motor will be progressively analyzed through two incremental models, as depicted in Figure 3. The first model, called $M_A$, considers only the leakage between both pole shoes. On the other hand, the second model, called $M_B$, also considers the leakage between pole shoe and armature, whose reluctances vary as a function of the armature position.

These reluctance networks describe the different flux paths generated by the electromagnetic device. A reluctance $\mathcal{R}_C$ in a material c is defined by an integral expression where the induction $B_c$ and cross-section $S_c$ vary linearly along the path of length $l_c$:

$$\mathcal{R}_C = \int_0^{l_c} \frac{dl}{\mu_c(B_c(l)) \times S_c(l)} \quad (1)$$

For leakage reluctances both length and cross-section are unknown as the flux tube in air widens unlike the case of flux concentration in iron. Only the air gap reluctances are assumed to be correctly modeled (through knowledge of geometric data). For iron reluctances, as explained in section I, the induction tangential component is not negligible, but unknown as it cannot be expressed as a function of magnetic flux.

For both models ($M_A$ & $M_B$), the electromagnetic torque is estimated using two analytical methods, one using Maxwell stress Tensor (TM), and other using the virtual work - also called magnetic COEnergy (COE), as presented in eq.2 and eq.3, respectively. Accordingly, four models are analyzed in this study ($M_A$ TM, $M_A$ COE, $M_B$ TM, $M_B$ COE). Noteworthy, the TM method uses the same assumptions as in (Merritt 1967; Urata 2007; Liu and Jiang 2016; Zhang et al. 2020; Meng et al. 2022), which neglect the tangential component of the air gap magnetic induction, and only the fluxes passing into the air gaps are considered for torque estimation.

$$TM: \Gamma = \frac{l_{la}}{\mu_0 S_g}\left(\Phi_{g1}^2 - \Phi_{g2}^2\right) \quad (2)$$

$$COE: : \Gamma = \frac{d}{d\theta}\left(\sum_i \left(\int_0^{V_{xi}} \Phi_{xi}(V_{xi})dV_{xi}\right)\right) \quad (3)$$

Furthermore, in both models, the permanent magnets are represented by a magnetic potential and reluctance denoted $V_m$ and $\mathcal{R}_m$, respectively.

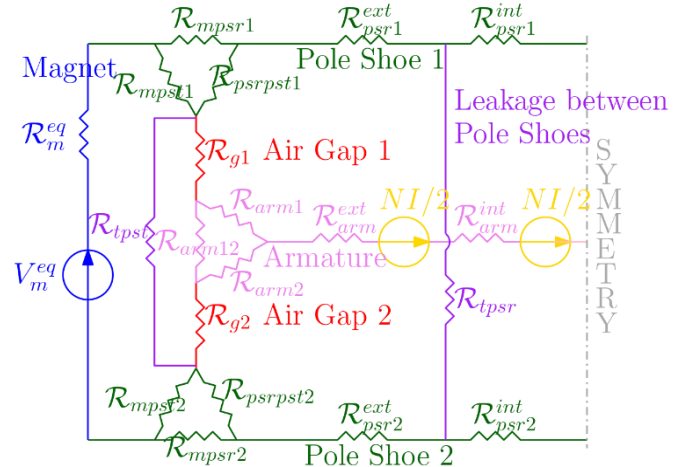

a. Reluctance model with leakage between pole shoes (Model $M_A$)

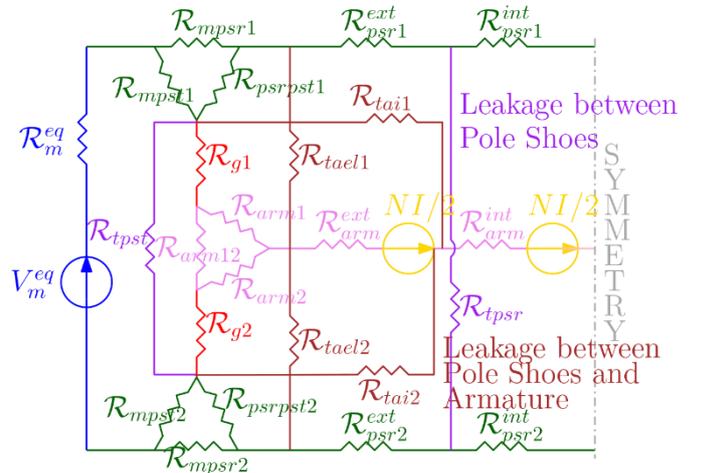

b. Reluctance model with leakage between pole shoes & pole shoes/armature (Model $M_B$)

*Figure 3. Different reluctance models of torque motor under study*

The first step of analytical model adjustment is done with a JMAG® simulation with iron infinite relative permeability, which allows to neglect every iron reluctance in the equivalent reluctance network and to get a faster determination of the leakage reluctance values. A set of data on torque versus current and angular position is collected from the JMAG® simulation at a fixed operating temperature, which are used as reference points for recalibrating leakage reluctances.

Both models have several unknowns to be identified via the reluctance adjustment process, such as $\mathcal{R}_{tpst}$, $\mathcal{R}_{tpsr}$, $\mathcal{R}_{tai1,2}$, $\mathcal{R}_{tael1,2}$, $k_m$. The coefficient $k_m$ represents the leakage located on the magnet ($V_m^{eq} = k_m V_m$ and $\mathcal{R}_m^{eq} = k_m \mathcal{R}_m$). The reluctances of leakage between pole shoes and armature $\mathcal{R}_{tai1,2}$, $\mathcal{R}_{tael1,2}$ are assumed to vary linearly with armature position. For $M_A$ method, 3 unknowns are adjusted to get the right torque instead of 7 unknowns for $M_B$ method.

The electromagnetic torque was assessed using these four models in our computational tool. Calculation results are validated by comparing it to JMAG® analysis results at different operating conditions (current and armature position). The Figure 4 compares the estimated torque by 4 methods versus the armature position at different current (33%In, 66%In and 100%In) with JMAG® for the same test conditions.

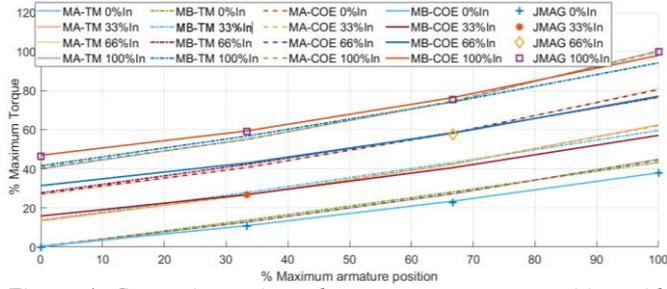

Figure 4. Comparing estimated torque vs armature position with JMAG results

| Modeling gap (%) | MA-TM | MB-TM | MA-COE | MB-COE |
|---|---|---|---|---|
| $I = I_n$  $\theta = 0°$ | 13.49 | 10.16 | 12.06 | 1.45 |
| $I = 0A$  $\theta = \theta_{max}$ | -18.66 | -14.46 | -17.68 | 0.32 |
| $I = I_n$  $\theta = \theta_{max}$ | -0.30 | 5.83 | 0.13 | 2.02 |

Tableau 1. Modeling gap between JMAG and different analytical models

As Figure 4 and Tableau 1 show, the best method to represent the electromagnetic torque is using the COEnergy (COE) with model $M_B$ that considers all leakages, where the maximum modeling discrepancy between this model and JMAG® is about 2% observed at $I_n$ & $\theta_{max}$. Therefore, this model will be considered as the advanced model, and consequently it will be used for the rest of the study.

### 4.2 Magnetic induction model

To represent more accurately the magnetic induction $B_c$ in the part $c$ (supposed uniform in the section $S_c$), it is needed to consider its normal component $B_c^n$ and its tangential component $B_c^t$. The normal component is usually represented by the well-known equation, eq.4:

$$\Phi_c = \iint \vec{B_c} \overrightarrow{dS_c} => B_c^n = \frac{\Phi_c}{S_c} \quad (4)$$

As mentioned previously, for the torque motor under study, the tangential component $B_c^t$ is not negligeable at the air gaps interfaces. It is therefore needed to estimate its value. The angle $\alpha_{iron}$ is introduced to link the normal and tangential component:

$$\alpha_{iron} = \tan^{-1}\left(\frac{B^t}{B^n}\right) \quad (5)$$

The angle $\alpha_{iron}$ is modelled for induction observed on air gaps (armature side and pole shoe side) in the iron by a 2nd order polynomial function dependent on the operating point. The tangential component of the induction is supposed to be zero at the other interfaces under study which means that only the calculation of $\mathcal{R}_{mpst1,2}$, $\mathcal{R}_{psrpst1,2}$, $\mathcal{R}_{pal1,2,12}$ are affected. The function depends on the inputs and outputs flux in the delta network under study, as shown in Figure 3 whose vary with operating point (angle, current).

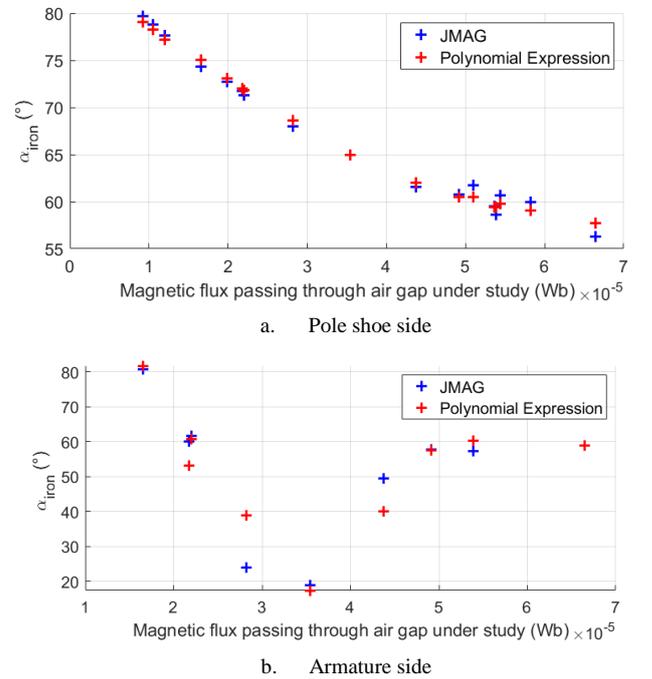

a. Pole shoe side

b. Armature side

Figure 5. Angle $\alpha_{iron}$ vs magnetic flux in air gap (a. Pole shoe side, b. Armature side)

The approximation of $\alpha_{iron}$ for both side of the air gaps are fairly accurate, with a maximum error of 1.4° for pole shoe side and 19.0° for armature side. A more significant error is visible on the armature side and could be reduced by modifying the polynomial degree or even providing another fitting function (e.g., multiplicative inverse). This error can lead to important gap between the modelled and the effective magnetic induction as total induction is expressed:

$$B_c = B_c^n \sqrt{1 + \tan^2(\alpha_{iron})} \quad (6)$$

The modeling error on $B_c$ becomes critical as $\alpha_{iron}$ evolves toward 90°.

## V APPLICATION OF DIFFERENT MATERIAL AND OPERATING CONDITIONS

The developed modeling tool allows to quickly assess the electromagnetic torque using other ferromagnetic materials (cheaper or with better physical or mechanical properties) and under different operating temperatures, which is very rewarding during the development phase.

### 5.1 Soft ferromagnetic material modification

The performance of torque motor under study that shown in section IV is for ferromagnetic material of (Fe-Ni), commercially known as Supra50®. This ferromagnetic material can be assimilated to an infinite relative permeability material in this case as Supra50® is well-known for its good properties when not saturated. As it is about 5 times more expensive than pure iron, it worth to evaluate the motor performance reduction using cheaper material. Therefore, the developed modeling tool is used to assess the torque for pure iron (ArmCo®) as a soft ferromagnetic material. Figure 6 compares the estimated torque using the developed model for pure iron, with JMAG® for both materials (pure iron & Supra50®).

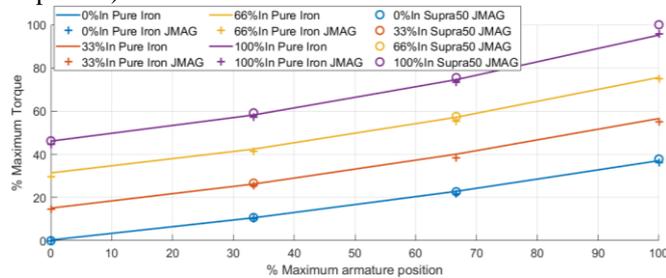

*Figure 6. Comparison torque vs armature position at diff currents for two soft materials for (developed model & JMAG®)*

As Figure 6 shows, the drop in the estimated torque for pure iron remains below 5% for the analytical model and for the JMAG® analysis. The analytical model tends to overestimate the torque, but the gap remains correct. Only 40s to 45s are needed to analyze the use of the new material, compared with a much longer time for the JMAG® simulation (for this motor with the chosen mesh, around 45min).

It should be noted that electromagnetic performance is assessed at steady state. The Supra50® is well known for its good dynamic performance with less iron loss and thus hysteresis which limits the benefits of using pure iron.

### 5.2 Operating temperature modification

All previous torque motor performance analysis were performed at room temperature of 20°C. As the magnetic properties of iron parts and permanent magnets decrease with increasing temperature, so it is interesting to evaluate the performance reduction at different temperatures, using the developed tool. It is assessed at two temperatures of 200°C and 350°C (maximum operating temperature given by the magnet manufacturers), as depicted in Figure 7. In this study, only the effect on magnetism is modelled without considering modification on electric and mechanic behavior of the torque motor.

The results of analytical model are quite close to those of JMAG®, which tends to validate the proposed approach. For instance, at 200°C, the maximum torque reduction, which is observed at nominal current and maximum angular position, is about 13% for analytical model and 10.7% for JMAG®, whereas it becomes higher at 350°C, as expected, where it is about 19.9% and 16.7% for analytical model and JMAG® simulation, respectively.

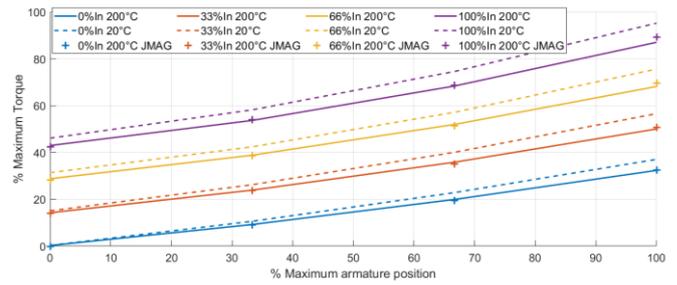

a. Torque vs armature position at 200°C & 20°C

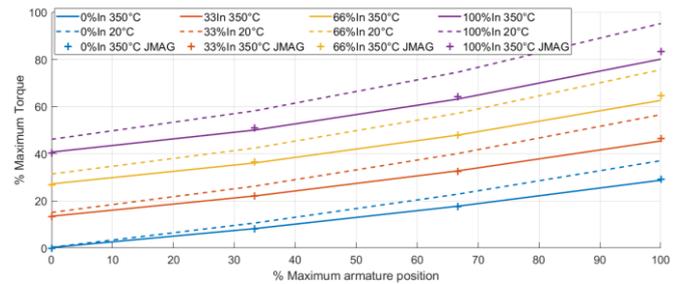

b. Torque vs armature position at 350°C & 20°C

*Figure 7. Comparison torque vs armature position at diff currents & for diff T° (a. at 200°C, b. at 350°C) vs. 20°C for (developed model & JMAG®)*

Another concern is about demagnetization of the magnet at high temperature, which can occur below a certain limit of magnetic induction. In our case, it is checked that there is no risk of demagnetization, as induction remains above this limit.

## VI CONCLUSION

In this paper, a new torque motor modeling tool is presented. It is based on an advanced reluctance model of the torque motor to reproduce a multivariable function that estimates the electromagnetic torque whatever the motor geometry, ferromagnetic materials properties, and operating conditions (current, armature position, temperature). This function can be integrated into any 1D multi-physics software to assess the effect of torque motor performance on the overall performance of servovalve system under different operating conditions.

The proposed advanced model is based on the reluctance network model that considers the permanent magnet reluctances, ferromagnetic materials reluctances, and the leakage reluctances (of permanent magnets, between the two stationary pole shoes, and between the pole shoe and the armature). Two incremental models are presented that progressively show the effect of different leakage reluctances on the torque motor performance, particularly this between the pole shoes and the armature. In addition, the Maxwell stress Tensor (TM) and the magnetic COEnergy method (COE) are used as analytical methods to calculate the electromagnetic torque for both models. Thus, 4 solutions have been analyzed and compared to FEM analysis results in JMAG® software. Noteworthily, the tangential component of iron magnetic induction is also considered in all models, contrary to other studies available in literature. Thus, the advanced model is the one that incorporates all leakage reluctances and utilizes the

COEnergy method, exhibiting superior performance when compared to FEM results.

The developed methodology enables a quick assessment of static electromagnetic performance at different servovalve operating temperature and using different ferromagnetic materials.

# NOTATIONS

Variables and parameters:

| | | |
|---|---|---|
| $T$ | Actuator operating temperature | °C |
| $M$ | Ferromagnetic alloys used | |
| N | Spire number by coil | turns |
| I | Applied current in coils | A |
| $\Gamma$ | Electromagnetic torque | N.m |
| $\theta$ | Angular position of armature | rad |
| $\mu$ | Permeability | H/m |
| $\mathcal{R}$ | Reluctance | H$^{-1}$ |
| $V$ | Magnetic potential drop | A |
| $\Phi$ | Magnetic flux | Wb |
| $l$ | Length | m |
| $S$ | Cross-section | m² |

Subscripts:

| | |
|---|---|
| $n$ | Nominal |
| $x, c$ | In random material or part |
| $i$ | Side of air gap |
| $m$ | Magnet part |
| $ps$ | Pole shoe part |
| $psr$ | Straight part of the pole shoe (above the coils) |
| $pst$ | Teeth part of the pole shoe (air gap side) |
| $arm$ | Armature part |
| $t$ | Leakages |
| $tpsr$ | Leakages between straight part on the pole shoe |
| $tpst$ | Leakages between both teeth on the pole shoe |
| $tael$ | Leakages on the outside of armature before coils |
| $tai$ | Leakages on the inside of armature in the center of the coils |
| $la$ | Lever arm |

Superscripts:

| | |
|---|---|
| 0 | Initial configuration |
| $j$ | New configuration |
| $n$ | Normal component |
| $t$ | Tangential component |
| $eq$ | Equivalent model |